\documentclass[english,english,aps,prd,amsmath,amssymb,superscriptaddress]{revtex4}
\usepackage[T1]{fontenc}
\usepackage[latin9]{inputenc}
\usepackage{amsmath}
\usepackage{amssymb}
\usepackage{esint}

\makeatletter
\@ifundefined{textcolor}{}
{%
 \definecolor{BLACK}{gray}{0}
 \definecolor{WHITE}{gray}{1}
 \definecolor{RED}{rgb}{1,0,0}
 \definecolor{GREEN}{rgb}{0,1,0}
 \definecolor{BLUE}{rgb}{0,0,1}
 \definecolor{CYAN}{cmyk}{1,0,0,0}
 \definecolor{MAGENTA}{cmyk}{0,1,0,0}
 \definecolor{YELLOW}{cmyk}{0,0,1,0}
}

\usepackage{slashed}\usepackage{babel}\usepackage{bbm}

\makeatother

\usepackage{babel}

\makeatother

\usepackage{babel}
\begin{document}

\title{The Cosmological Constant Problem and Quantum Spacetime Reference
Frame}

\author{M.J.Luo}

\email{mjluo@ujs.edu.cn}

\address{Department of Physics, Jiangsu University, Zhenjiang 212013, People's
Republic of China}
\begin{abstract}
We generalize the idea of quantum clock time to quantum spacetime
reference frame via physical realization of a reference system by
quantum rulers and clocks. Omitting the internal degrees of freedom
(such as spins) of the physical rulers and clocks, only considering
their metric properties, the spacetime reference frame is described
by a bosonic non-linear sigma model (NLSM). We study the quantum behavior
of the system under approximations, and obtain (1) a cosmological
constant valued $(2/\pi)\rho_{c0}$ ($\rho_{c0}$ the critical density
at near current epoch) which is very close to the observations; (2)
an effective Einstein-Hilbert term; (3) the ratio of variance to mean-squared
of spacetime interval tends to a universal constant $2/\pi$ in the
infrared region. This effect is testable by observing a linear dependence
between the inherent quantum variance and mean-squared of the redshifts
from cosmic distant spectral lines. The proportionality is expected
to be the observed percentage of the dark energy. The equivalence
principle is also generalized to the quantum level. 
\end{abstract}
\maketitle

\section{Introduction}

Reference frame is one of the most fundamental notions in physics.
When a measurement in physics is performed or described, a reference
frame has always been explicitly or implicitly used. As ordinarily
formulated, reference frame idealizationally uses the rulers and clocks
to label the spacetime for simplicity, which have well-defined values
of coordinates and are considered most perfect, absolute, classical,
and external. This fundamentally classical notion of reference frame
has been using in almost all area of physics including today's textbook
quantum physics, although quantum mechanics has been discovered for
a century. The quantum mechanics tells us that all measuring devices
are subject to some level of quantum fluctuations, certainly applying
to the rulers and clocks, namely the spacetime. Such idealizational
treatment works well in quantum mechanics and quantum field theory
when the equations of them cast in terms of the variables that are
really measured by physical rulers and clocks in ordinary laboratory.
This is, to a large extent, due to the fact that gravitational effects
are not seriously taken into account in most laboratory experiments.
Since according to the standard theory of gravity, the general relativity,
the spacetime is dynamical and relational. It is as expected, when
the quantum mechanics is applied to the cosmology which is gravity
dominated, severe difficulty arises: the cosmological constant problem,
see for instance \cite{RevModPhys.61.1,2008GReGr..40..607B,2012CRPhy..13..566M}
and references therein.

Along this line of thinking, treating the concept of a reference frame
in quantum theory may be the key to the cosmological constant problem.
The earlier publications \cite{2014NuPhB.884..344L,2015JHEP...06..063L}
have proposed a possible solution by replacing the idealized parameter
time in textbook quantum theory by a quantum dynamical variable playing
the role of a physical clock time. The papers obtain a cosmological
constant having not only a correct order but also a percentage $\Omega_{\Lambda}=2/\pi\approx0.64$
which is very close to current observational value \cite{2013arXiv1303.5062P}.
However, such solution can only be regarded as incomplete, since although
time has been treated quantum mechanically, in those papers, the spatial
coordinates are still treated as classical external parameters free
from quantum fluctuations. The space and time measurements are closely
related to each other and must be treated on an equal footing. In
this work, we shall generalize the discussion of quantum clock time
in refs.\cite{2014NuPhB.884..344L,2015JHEP...06..063L} to a more
general framework: quantum spacetime reference frame. It is a years
old idea (see for instance \cite{1984PhRvD..30..368A,rovelli1991quantum,dickson2004view,angelo2011physics}
and references therein), but to the best of our knowledge, there is
no literature or discussion yet connecting such idea to the cosmological
constant problem, the goal of the paper is to show their deep relation.
We take the natural unit $\hbar=c=1$ for convention in the paper.

\section{Quantum Physical System Relative to a Quantum Reference Frame}

When a reference frame (rulers, clocks or other measuring devices),
to which a to-be-studied quantum physical system is relative, are
inherent quantum, they can not be just ignored from a complete quantum
description. In a complete quantum treatment, a set of states of the
to-be-studied quantum physical system (denoted by P) together with
the quantum reference frame system (denoted by R) are described by
a Hilbert space $\mathcal{H}$ being a direct product of both Hilbert
spaces,
\begin{equation}
\mathcal{H}=\mathcal{H}_{P}\otimes\mathcal{H}_{R}.
\end{equation}

However, it does not mean that the state vector of the whole system
is simply a direct product of vector in each Hilbert space. In most
cases, it is an entangled state, the reason is as follows. Generally,
in a first step of performing a physical measurement, instrument calibration
must be carried out firstly. A good experimental calibration establishes
a one-to-one correlation between a state of the to-be-studied quantum
physical system $\left|P\right\rangle _{i}$ and a state of the measuring
device $\left|R\right\rangle _{j}$. If the calibration is well done,
such step introduces a complete bases $\left|P\right\rangle _{i}\otimes\left|R\right\rangle _{j}$
to expand a state $\left|\Psi\right\rangle $ in the whole Hilbert
space $\mathcal{H}$,
\begin{equation}
\left|\Psi\right\rangle =\sum_{i,j}\alpha_{ij}\left|P\right\rangle _{i}\otimes\left|R\right\rangle _{j}.\label{eq:entangled-state}
\end{equation}
In the second step, physicist let these two systems evolve independently
and observe the output of the measuring device. The physics of the
individual measuring device is assumed well understood, having complete
knowledge of the Hilbert space $\mathcal{H}_{R}$, and a state in
it can be expanded by a complete bases $\left|R\right\rangle =\sum_{j}\beta_{j}\left|R\right\rangle _{j}$.
Then physicist uses the information of calibration Eq.(\ref{eq:entangled-state})
to judge what is the state of P when the output of the measuring device
being in one of the state $\left|R\right\rangle _{j}$. The expansion
coefficient $\alpha_{ij}$ measures the amplitude of the measuring
device being in the state $\left|R\right\rangle _{j}$, and meanwhile,
the to-be-studied system P being in the state $\left|P\right\rangle _{i}$.
The $\left|\alpha_{ij}\right|^{2}$ measures the joint probability
that P and R are in the states $\left|P\right\rangle _{i}$ and $\left|R\right\rangle _{j}$,
respectively.

\subsection{Relational Interpretation}

The entangled state has a general property that the amplitude $\alpha_{ij}$
can not be factorized into a product of each amplitude of $\left|P\right\rangle $,
$\left|R\right\rangle $, namely, for all amplitudes $\gamma_{i}$,
$\beta_{j}$, defined by $\left|P\right\rangle =\sum_{i}\gamma_{i}\left|P\right\rangle _{i}$
and $\left|R\right\rangle =\sum_{j}\beta_{j}\left|R\right\rangle _{j}$,
we always have $\alpha_{ij}\neq\gamma_{i}\beta_{j}$. This property
has important physical implications.

A physical meaningful probability in a measurement is the probability
of state $\left|P\right\rangle _{i}$ given the condition that the
measuring device is observed in the state $\left|R\right\rangle _{j}$.
The conditional probability $\mathrm{Pro}(P_{i}|R_{j})$ is defined
by
\begin{equation}
\mathrm{Pro}(P_{i}|R_{j})\equiv\frac{\mathrm{Pro}(P_{i}\cap R_{j})}{\mathrm{Pro}(R_{j})}=\frac{\left|\alpha_{ij}\right|^{2}}{\left|\beta_{j}\right|^{2}},
\end{equation}
 in which the $\mathrm{Pro}(P_{i}\cap R_{j})$ stands for the joint
probability $\left|\alpha_{ij}\right|^{2}$. The $\mathrm{Pro}(R_{j})$,
which is given by $\left|\beta_{j}\right|^{2}$, represents the probability
of state $\left|R\right\rangle _{j}$ when the measuring device independently
evolves in the second step. Since $\alpha_{ij}\neq\gamma_{i}\beta_{j}$,
the conditional probability that P is in $\left|P\right\rangle _{i}$
given R being in $\left|R\right\rangle _{j}$ does not equal to the
probability that P is individually in the state $\left|P\right\rangle _{i}$,
i.e. $\mathrm{Pro}(P_{i}|R_{j})\neq\mathrm{Pro}(P_{i})$. In this
sense, the probability of $\left|P\right\rangle _{i}$ is affected
by the probability distribution of $\left|R\right\rangle _{j}$. The
state $\left|P\right\rangle _{i}$ of the to-be-studied system has
no individual absolute meaning, it makes sense only relative to the
state $\left|R\right\rangle _{j}$ of the measuring device as a reference
frame \cite{PhysRevD.27.2885,2014PhRvA..89e2122M}, namely, the state
is relational \cite{1996IJTP...35.1637R,rovelli2004quantum}.

The entangled state is purely a quantum state, having no classical
correspondence. Only when the measuring devices are treated semi-classically,
i.e. a delta distribution for $\left|R\right\rangle _{j}$, the sign
of inequality becomes an equal sign, in this limit it recovers the
absolute probability interpretation of the textbook quantum mechanics.
For instance, in general the distribution for $\left|R\right\rangle _{j}$
has a finite width due to quantum fluctuation, one can prove that
it is equivalent to a textbook wavefunction of P with a smeared variable
R \cite{phillips2000vacuum}.

In this paper, we argue that when a quantum theory is completely formulated
in terms of the states entangling a to-be-studied quantum physical
system with a quantum reference frame or quantum measuring devices,
and interpreted it in a relational manner, the quantum theory is able
to accommodate the spirit of relativity, leading to a consistent theory
of quantum spacetime reference frame.

\subsection{Identical Particles Model of Spacetime Reference Frame}

To take a further look into the Hilbert space of the reference frame
$\mathcal{H}_{R}$, let us considering an operative setup to realize
the physical rulers and clocks at the quantum level. Suppose we have
$D$ free identical scalar particles, living in a $d$-dimensional
background $x_{a}$, $(a=0,1,2,...,d-1)$. Omitting the internal degrees
of freedom of them such as spins, let $X_{\mu}(x)$, $(\mu=0,1,2,...,D-1)$
represent scalar fields of these identical particles, with dimension
$[\mathrm{mass}]^{-1}$. As ordinarily described, the frame $x$ can
be interpreted as the coordinates according to the walls and clock
of a laboratory. Assuming that these identical scalar fields evolve
independently in the laboratory, the resulting action is the summation
of each particle, namely, the action is separable,
\begin{equation}
S_{R}[X]=\sum_{\mu}^{D}S_{id}[X_{\mu}].\label{eq:action-frame}
\end{equation}
The action is separable means that the Hilbert space of the identical
particles system is a direct product of each particles $\mathcal{H}_{R}=\otimes_{\mu}^{D}\mathcal{H}_{X_{\mu}}$,
however, as is shown below, for the nontrivial interchange properties,
the state of the identical particles or equivalently the spacetime
reference frame is also an entangled state of each particle, instead
of a simply direct product. For each identical particles, $S_{id}[X_{\mu}]$
is a standard free scalar field action formulated in a homogeneous
and flat frame with respect to the laboratory walls and clock 
\begin{equation}
S_{id}[X_{\mu}]=\frac{\lambda}{2}\int d^{4}x\partial_{a}X_{\mu}\partial_{a}X_{\mu},\quad(\mu=0,1,2,3)\label{eq:free-scalar-reference-action}
\end{equation}
in which $\partial_{a}=\frac{\partial}{\partial x_{a}}$, and $\lambda$
is a constant having dimension $[\mathrm{mass}]^{d}$ and related
to a ``rest mass'' scale (with respect to the frame $x$) of the
identical scalar particles, since we have used the laboratory wall
interpretation to $x_{a}$ so here $d=4$. 

In a range of ordinary laboratory scale, the walls and clock of the
laboratory can be used to orient, align and order the beams of these
scalar particles at high precision, in this sense, these identical
quantum scalar free fields can be interpreted as physical rulers and
clock at the quantum level. For example, the identical scalar fields
oriented as $X_{1,2,3}=X,Y,Z$ can be aligned with a reference to
the $x,y,z$-directions of the walls of the laboratory, respectively.
One could visualize them as local quantum vibrations or oscillations
placed on the lattice of $x,y,z$, these identical particles can be
seen as rulers since distances are able be measured by counting their
phase changes of the local vibrations if events trigger the counting.
In this sense they play the roles of state-triggers labeling where
the event happens. For the same consideration, a scalar field denoted
as $X_{0}=T$ on the lattice can be used to play the role of a small
pendulum clock labeling the causal order of the events, i.e. when
the event happens. As a result, an event denoted by state $\left|P\right\rangle _{i}$
is entangled with a particular configuration of the identical particles
system it triggers 
\begin{equation}
\left|R\right\rangle _{j}=\left|X_{0},X_{1},X_{2},X_{3}\right\rangle ,
\end{equation}
according to the entanglement Eq.(\ref{eq:entangled-state}). So we
could say that a state of the identical particles as a standard reference
system labels where and when a quantum event happens, playing the
role of a spacetime reference frame.

For the bosonic statistics of these four identical particles, the
interchange between them does not change the state $\left|R\right\rangle _{j}$.
In one case, because the role of the vibrations on the lattice as
ruler or clock is just a convention, if we interchange $X_{0}$ and
$X_{i},\;(i=1,2,3)$ the system obviously does not change. In another
case, if we interchange two rulers' fields, for instance, $X$ and
$Y$, the frame changes from left-hand to right-hand, but we have
the state unchanged, i.e. $\left|X_{0},X_{1},X_{2},X_{3}\right\rangle =\left|X_{0},X_{2},X_{1},X_{3}\right\rangle $,
which implies a reflection symmetry (the parity symmetry) of spacetime
coordinates. In general, the configuration space of the identical
particles is $\mathbb{R}^{4}/S_{4}$, where $S_{4}$ is the permutation
symmetry of the identical particles representing the discrete symmetries
of spacetime and/or their certain products. 

A practical example for the identical scalar particles model of a
spacetime reference frame is the multi-wire chamber. The beams of
the scalar fields used in the model can be considered as free electron
fields in the array of multi-wire, which signal the coordinates of
an event by an impulse at the output. As it is pointed out, the electron
signal triggered by the event is inescapably quantum, more precisely,
the electron obeys the quantum uncertainty principle.

When the scale of an experimental measurement is much larger than
the scale of an ordinary laboratory, the laboratory wall interpretation
becomes no longer valid, since the distance can not be measured directly
as the previous situation in the laboratory. What we could use to
measure spacetime coordinates is inevitably only by using the identical
particles being our rulers and clock. For example, in the situation
of a cosmic observation, the scalar particles used in the model can
be instead considered as free photons. The information of distance
of an event can only be extracted from observations of, e.g. luminosity
and frequency/redshift of distant spectral lines.

Here the notion of metric is in general non-trivial in the situation
that the spacetime coordinates is operationally measured by the physical
fields as rulers and clock. Since a realistic geometry measured via
the free fields in general does not necessarily match the geometry
which is expected beyond the laboratory. For example, a ruler or clock
elsewhere measured by a field $X_{i}$ or $X_{0}$ requires certain
techniques to compare or synchronize to the laboratory ruler $x_{a}$
or clock $x_{0}$, regardless that the laboratory rulers and clock
are realistic or extrapolative or complete imaginary. Briefly speaking,
the identical particles propagates in a general curved spacetime measured
by themselves. By using mathematical method, the metric of the actual
spacetime can be given by a vierbein, which makes a comparison between
the physical coordinates $X_{\mu}$ and the extrapolated absolute
coordinates $x_{a}$, $e_{\mu}^{a}\equiv\partial X_{\mu}/\partial x_{a}$.
The metric having dimension $[\mathrm{mass}]^{0}$ is defined by 
\begin{equation}
g_{\mu\nu}(x)=e_{\mu}^{a}e_{\nu}^{a}=\frac{\partial X_{\mu}(x)}{\partial x_{a}}\frac{\partial X_{\nu}(x)}{\partial x_{a}},\label{eq:metric}
\end{equation}
in which the extrapolated laboratory wall frame $x$ is assumed to
be homogeneous and flat as well. The assumption is considered having
no impact on the physical result since the physics does not depend
on the choice of frame, which can seen more clearly later. Precisely
speaking, we can not align and order the beams of these identical
free scalar particles according to the walls and clock of the laboratory
in prior, but in general according to the metric $g_{\mu\nu}=\left(g^{\mu\nu}\right)^{-1}$,
which is practically measured. In this situation, the action of identical
particles is generalized to
\begin{equation}
S_{R}[X]=\frac{\lambda}{2}\int d^{d}x\sum_{\mu}^{D}\partial_{a}X^{\mu}\partial_{a}X_{\mu}=\frac{\lambda}{2}\int d^{d}x\sum_{\mu,\nu}^{D}g_{\mu\nu}\partial_{a}X^{\mu}\partial_{a}X^{\nu},\label{eq:NLSM}
\end{equation}
where the indices are raised or lowered by the metric, e.g. $X^{\mu}=g^{\mu\nu}X_{\nu}$.
In the action the configuration space of the identical particles locally
is generalized from $\mathbb{R}^{4}$ to a more general Riemannian
space $(\mathbb{M}^{4},g)$ with non-trivial metric, and we do not
pre-assume the value of dimension $d$ from the beginning whose value
will be discussed in detail later.

For the reason that a $D=4$ curved manifold can be isometrically
embedded in a linear space of $D+1$ dimensions, one can also start
from a linear theory without prior geometric notion such as metric
of the target manifold
\begin{equation}
S[\phi]=\int d^{d}x\left[\frac{\lambda}{2}\sum_{\mu=1}^{D+1}\partial_{a}\phi^{\mu}\partial_{a}\phi^{\mu}+\frac{1}{2}i\xi(\sum_{\mu=1}^{D+1}\lambda\phi^{\mu}\phi^{\mu}-M^{d-2})\right],\label{eq:NLSM-1}
\end{equation}
in which $\xi$ is a Lagrangian multiplier that imposes the constraint
$\sum_{\mu=1}^{D+1}\lambda\phi^{\mu}\phi^{\mu}=M^{d-2}$, where $M$
is considered a fundamental energy scale. When we functionally integrate
out $\xi$ with the identification $\phi=(\delta X^{0},\delta X^{1},\delta X^{2},\delta X^{3},\sigma)$,
Eq.(\ref{eq:NLSM-1}) can be formulated in the form of Eq.(\ref{eq:NLSM})
\begin{equation}
S[\delta X]=\frac{\lambda}{2}\int d^{d}x\sum_{\mu,\nu}^{D}\left(\eta_{\mu\nu}+\frac{\delta X_{\mu}\delta X_{\nu}}{\sigma^{2}}\right)\partial_{a}(\delta X^{\mu})\partial_{a}(\delta X^{\nu}),
\end{equation}
with an induced metric 
\begin{equation}
\tilde{g}_{\mu\nu}[X]=\eta_{\mu\nu}+\frac{\delta X_{\mu}\delta X_{\nu}}{\sigma^{2}}.\label{eq:metric-linear}
\end{equation}
 The induced metric Eq.(\ref{eq:metric-linear}) is equivalent to
the expectation value of the metric Eq.(\ref{eq:metric}), since according
to the equivalence principle we can always choose a local inertial
reference frame such that the spacetime $X_{\mu}$ seems to be locally
the flat parameter background $x$ (the laboratory wall frame), if
we expand $X_{\mu}$ in Eq.(\ref{eq:metric}) around the background
$x$ with fluctuations $X_{\mu}(x)=x_{\mu}+\delta X_{\mu}(x)$ for
any chosen point $x$, taking $\langle\delta X_{\mu}\rangle=0$ while
$\langle\delta X_{\mu}\delta X_{\nu}\rangle\neq0$, and considering
field value $\sigma$ measures the proper length $\left(\sum_{a}\delta x_{a}^{2}\right)^{1/2}$
of the parameter background. Furthermore, as the proper length of
the parameter background hardly vary with spacetime, the $\sigma$
field is therefore too massive to have any fluctuations and excited
states, the $\sigma$ field is merely an unobserved auxiliary field,
what we used to measure the realistic spacetime are those identical
particles $X_{\mu}$. 

For historic reasons, the action Eq.(\ref{eq:NLSM}) is known as the
non-linear sigma model (NLSM) \cite{gell1960axial,ketov2000quantum,zinn2002quantum}
having many applications in particle physics and condensed matter
physics. The idea of using a NLSM to describe dynamical coordinates
of spacetime has also been suggested in earlier works e.g.\cite{Omero1980Generalized,Gell1985Dimensional,2006PhRvD..74f4018G,Percacci2014Geometry},
our work goes further than them and tries to reveal a close relation
between the idea and the cosmological constant in quantum gravity. 

The NLSM action Eq.(\ref{eq:NLSM}) maps a d-dimensional flat absolute
parameter background $x_{a}$ into a $D$-dimensional target Riemannian
manifold $X_{\mu}$ which is the configuration space of the identical
particles. The target manifold is invariant under local $O(D)$ rotation
symmetry, $X^{\mu}\rightarrow X^{\mu\prime}=\Lambda_{\nu}^{\mu}X^{\nu}$.
Such compact symmetry can be easily transformed into noncompact $O(1,D-1)$
symmetry by an unimportant phase reversal (a $\pi$-phase shift) redefinition
to the clock field $X^{0}$, i.e. $X^{0}\rightarrow iX^{0}$ when
we initialize the field. The $O(D)$ or $O(1,D-1)$ symmetry can be
interpreted as the Lorentz symmetry of the spacetime reference frame.

An important point to be emphasized here is that in the rest of the
paper, the absolute parameter background $x_{a}$ previously interpreted
as an extrapolative absolute laboratory wall frame will have nothing
to do with the realistic spacetime. A parameter background is necessary
for a theoretical description of a quantum fields theory, but they
are not necessarily interpreted as the physical spacetime, the realistic
physical spacetime is what we measure from the identical scalar fields
$X^{\mu}$ being the standard reference system. This point is very
important in resolving the cosmological constant problem which will
be shown in subsection-D of the section III.

Note that the classical Lagrangian of NLSM is formally proportional
to $g_{\mu\nu}g^{\mu\nu}=D$, it is the dimension of the target manifold
which is an invariant under the parameter background $x$-coordinates
transformations, so the NLSM is parameter background independent which
is the reason we could choose a flat background without loss of any
physical generality. Furthermore, the action is also invariant under
an arbitrary differentiable $X^{\mu}$-coordinates transformation
of the target manifold. In this sense, a system relative to such spacetime
reference frame is connected to a theory of gravity on the target
manifold, their relationship will be shown in the next section.

A choice of a reference $\mathcal{H}_{R}$ is mathematically equivalent
to choose a complete set of bases to formulate the whole Hilbert space
$\mathcal{H}=\mathcal{H}_{P}\otimes\mathcal{H}_{R}$. More precisely,
a choice of a spacetime reference frame explicitly solves the diffeomorphism
constraints \cite{rovelli1991quantum} and the Wheeler-DeWitt equation.
The constraints say that physics dose not depend on the choice of
reference frame, so if you choose one frame, a physical state must
be a summation over all possible choices. In this sense, it again
indicates that a physical state in $\mathcal{H}$ is not a simple
direct product state, it is an entangled state as Eq.(\ref{eq:entangled-state})
summing over all possible direct product states.

\subsection{Semi-Classical Approximation}

In this subsection we will take a first look at the whole system consisting
of a to-be-studied quantum physical system P and a quantum spacetime
reference frame R which P relative to. After the calibration, the
entanglement between P and R is developed. Then P and R evolve independently,
and we assume that they do not interact, so the total action can be
separably written as
\begin{equation}
S[\varphi,X]=S_{P}[\varphi]+S_{R}[X].\label{eq:total-action}
\end{equation}
 Without loss of generality, we consider the to-be-studied system
is a scalar fields theory formulated in the same laboratory wall frame
as that formulating the spacetime reference system Eq.(\ref{eq:NLSM}),
\begin{equation}
S_{P}[\varphi]=\int d^{d}x\left(\frac{1}{2}\partial_{a}\varphi\partial_{a}\varphi-V(\varphi)\right).\label{eq:scalar-field-action}
\end{equation}
So that the reference frame R and the to-be-studied system P share
the same laboratory coordinates $x$, the total action is then given
by
\begin{equation}
S[\varphi,X]=\int d^{d}x\left[\frac{1}{2}\partial_{a}\varphi\partial_{a}\varphi-V(\varphi)+\frac{\lambda}{2}g_{\mu\nu}\partial_{a}X^{\mu}\partial_{a}X^{\nu}\right].\label{eq:Sp+Sr}
\end{equation}
In a semi-classical approximation in which the fields $X_{\mu}$ can
be seen as c-numbers, the action can be rewritten as
\begin{equation}
S_{eff}[\varphi]=\int d^{4}X\sqrt{\mathrm{det}g}\left[\frac{1}{4}\langle g_{\mu\nu}\partial_{a}X^{\mu}\partial_{a}X^{\nu}\rangle\left(\frac{1}{2}g^{\mu\nu}\frac{\delta\varphi}{\delta X^{\mu}}\frac{\delta\varphi}{\delta X^{\nu}}+2\lambda\right)-V(\varphi)\right],\label{eq:semi-classical-action-I}
\end{equation}
where we have used $g_{\mu\nu}g^{\mu\nu}=D=4$ and $\sqrt{\mathrm{det}g}=\left\Vert \frac{\partial x_{a}}{\partial X^{\mu}}\right\Vert $
is the Jacobian determinant transforming the integration variable
from $x$ to $X$. The Jacobian matrix is a square matrix, the property
requires the dimension of the space $x$ equaling to the number of
the identical particles $X$, so $d=D=4$. It must be emphasized that
this statement is true only in the semi-classical approximation, it
is not necessarily true beyond the approximation as will be shown
in the next section when we study its quantum behavior. Since $\langle g^{\mu\nu}\rangle=\langle\partial_{a}X^{\mu}\partial_{a}X^{\nu}\rangle$,
so $\frac{1}{4}\langle g_{\mu\nu}\partial_{a}X^{\mu}\partial_{a}X^{\nu}\rangle=1$.
We obtain
\begin{equation}
S_{eff}[\varphi]=\int d^{4}X\sqrt{\mathrm{det}g}\left[\frac{1}{2}g^{\mu\nu}\frac{\delta\varphi}{\delta X^{\mu}}\frac{\delta\varphi}{\delta X^{\nu}}-V(\varphi)+2\lambda\right].\label{eq:semi-classical-action-II}
\end{equation}

The effective action obtained from the semi-classical approximation
has a straightforward interpretation: the to-be-studied quantum fields
system of $\varphi$ is relative to the spacetime reference frame
$X^{\mu}$ semi-classically treated. The equation Eq.(\ref{eq:semi-classical-action-II})
is similar with Eq.(\ref{eq:scalar-field-action}) up to a constant
shift, just formally the derivative $\frac{\partial}{\partial x}$
is replaced by a functional derivative $\frac{\delta}{\delta X}$,
and the function $\varphi(x)$ with respect to $x$ is replaced by
a functional $\varphi[X]$ with respect to $X$.

Furthermore, although parameter background space $x$ is flat, when
the target manifold is curved, the effective action describes such
a theory that field or even quantized field $\varphi$ is in a curved
spacetime. So we prove the equivalence between the semi-classical
theory of the spacetime reference frame and the theories of (quantum)
fields in curved spacetime. Thus, it is reasonable to expect that
the theory could recover the existing results of quantum fields theories
in curved spacetime, for instance, the Hawking's radiation of a black
hole.

However, only a semi-classical treatment of the spacetime reference
frame is not enough, where the spacetime is still fixed, the quantum
dynamics of the spacetime reference frame must be considered. We will
show in the next section that an effective Einstein's theory of gravity
and a correct cosmological constant naturally arise from the quantum
behavior of the spacetime reference frame.

\section{Quantum Behavior}

In this section, we will study the quantum dynamics of the spacetime
reference frame defined in previous section. Let us recall the NLSM
in Eq.(\ref{eq:NLSM}), which maps a $d$ dimensional flat homogeneous
parameter background $x_{a},(a=0,1,2,...,d-1)$ into a $D$-dimensional
target manifold $X^{\mu},(\mu=0,1,2,...,D-1)$. Here we interpret
the target manifold coordinates $X^{\mu}$ as scalar fields describing
the coordinates of the realistic spacetime, so $D\equiv4$ is considered
fixed. However, it is not necessary to fix the dimension $d$ of parameter
background from the beginning, since first it is unobservable in the
theory and second it is known that $d$ runs as renormalization flows,
the discrepancy is known as the anomalous dimension. Therefore here
the parameter background space has nothing to do with our realistic
spacetime any more in the theory, and $d$ is considered varying with
the scale of renormalization. The action of the NLSM is
\begin{equation}
S_{NLSM}=\frac{\lambda}{2}\int d^{d}xg_{\mu\nu}[X]\partial_{a}X^{\mu}\partial_{a}X^{\nu},\label{eq:NLSM-UV}
\end{equation}
where $g_{\mu\nu}[X]$ is a positive dimensionless metric of the target
manifold, and since coordinates $x_{a}$ and $X^{\mu}$ have dimensions
$[\mathrm{mass}]^{-1}$, the constant $\lambda$ has dimension $[\mathrm{mass}]^{d}$.

\subsection{Renormalization}

Most of the quantum behaviors of a system encode in its renormalization.
For the sake of making the treatment focus on the cosmological constant,
we only discuss a global renormalization function $Z(k)=1+\delta_{Z}(k)$,
which rescales each field isotropically,
\begin{equation}
X^{\mu}\rightarrow X_{re}^{\mu}=Z^{1/2}(k)X^{\mu}.\label{eq:definition-Z}
\end{equation}
The parameter $k$, with dimension $[\mathrm{mass}]^{1}$, introduced
by hand measures a cutoff of the Fourier component of the dynamical
spacetime fields $X^{\mu}$. So here $k$ replaces the absolute parameter
background $x$, playing the role of a renormalization evolution parameter.
Then the action in a ``Wilsonian'' sense is effective defined at
the cutoff $k$,
\begin{equation}
S_{k}=\frac{\lambda}{2}\int d^{d}xZg_{\mu\nu}\partial_{a}X^{\mu}\partial_{a}X^{\nu}.\label{eq:renormalized-action}
\end{equation}

It is worth emphasizing that the global renormalization function $Z$
can also be interpreted as a renormalization to the inverse coupling
constant $\lambda$ or even the geometric measure of the base space
$d^{d}x$ while keeping $X^{\mu}$ fixed, they have completely different
physical interpretations but in our discussions they are mathematically
equivalent. For different conveniences and purposes, in the following
discussions different interpretations are used. 

Recall that, in the semi-classical approximation, the dimensions of
the parameter background are identical with that of the target manifold,
i.e. $d=D\equiv4$. Although it seems that the NLSM in $d=4$ is perturbative
non-renormalizable by power counting, at the non-perturbative level
it is shown \cite{codello2009fixed} that a $d=4$ NLSM has a non-trivial
(non-Gaussian) UV fixed point (i.e. $k\rightarrow\infty$). In this
sense, this theory of quantum spacetime is ``asymptotically safe''\cite{weinberg1979ultraviolet},
which will be seen later by noting that $d$ seems effectively run
to a lower 2-dimension at $k\rightarrow\infty$. Therefore, the fields
$X^{\mu}$ and $\lambda$ in Eq.(\ref{eq:NLSM-UV}) can be seen as
bare values defined at the UV fixed point, at which the renormalization
condition is given by
\begin{equation}
\lim_{k\rightarrow\infty}Z(k)=1,\quad\lim_{k\rightarrow\infty}\delta_{Z}(k)=0.\label{eq:renormalization-condition}
\end{equation}
 In order to find the physics at IR, the question we want to ask is:
from this initial renormalization condition at UV, what value does
it take when the renormalization group flows to IR ($k\rightarrow0$)?

The answer to the question for the $d=4$ NLSM has been studied in
literature by perturbation theory when the value of $\lambda$ is
large \cite{codello2009fixed} (the coupling of the NLSM being the
inverse of $\lambda$). Fortunately in our theory the bare value $\lambda$
defined at UV is indeed large. It is expected of order $\lambda\sim\mathcal{O}(R_{*}k_{UV}^{d-2})$
\cite{codello2009fixed}, where $R_{*}$ is a positive Ricci scalar
curvature of the target manifold at a fixed point and $k_{UV}$ is
a constant relating to a UV scale much larger than the scale $k$
we are interested in, i.e. $k_{UV}\gg k$.

In this situation, a large $\lambda$ corresponds to a small coupling,
the perturbation calculation of $\delta_{Z}$ is reliable, which at
one loop is given by
\begin{equation}
\delta_{Z}(k)-\delta_{Z}(0)=\frac{1}{2}\frac{R}{\lambda D}\int_{0\leq\left|p\right|<k}\frac{d^{d}p}{(2\pi)^{d}}\frac{i}{p^{2}}=\frac{1}{(4\pi)^{d/2}\Gamma(\frac{d}{2})(d-2)}\frac{R}{\lambda D}k^{d-2},\label{eq:general-delta}
\end{equation}
where $R$ is a positive induced Ricci scalar curvature of the target
manifold with dimension $[\mathrm{mass}]^{2}$. By using $d=4+\epsilon$
expansion and the minimal subtraction scheme, we obtain a regularized
result
\begin{equation}
\delta_{Z}(k)=C_{d}\frac{R}{\lambda D}k^{d-2}+C,
\end{equation}
where $C_{d}^{-1}=(4\pi)^{d/2}\Gamma(d/2+1)$ and $C$ represents
an integral constant taking the value $\delta_{Z}(0)$ to be determined
by the initial renormalization condition Eq.(\ref{eq:renormalization-condition}).
Thus for $d=4$ the renormalization function in the IR region behaves
like
\begin{equation}
\delta_{Z}(k)=\frac{1}{128\pi^{2}}\frac{R}{\lambda}k^{2}+C,\quad(\mathrm{for}\;\mathrm{small\;}k).\label{eq:delta-IR-1}
\end{equation}

To determine the integral constant $C$ we need to apply the initial
renormalization condition defined at UV. However, the function Eq.(\ref{eq:delta-IR-1})
is only valid at IR, obvious it diverges in the limit $k\rightarrow\infty$.
To a large extent, the difficulty can be attributed to the fact that
such behavior of $\delta_{Z}$ will completely change in the UV region
as a consequence of the running of the effective dimension $d$ of
the base space. In fact, the scale dependent anomalous dimension $\eta(k)$
always go with dimension $d$ and produces a scale dependent effective
$\epsilon$-parameter (expansion) of the dimensional continuation,
that is to say, everywhere that appears the dimension $d$ could be
approximately replaced by $d_{eff}=d-\eta(k)$. Since the anomalous
dimension equals to $\eta(k\rightarrow\infty)=d-2$ at the non-trivial
UV fixed point \cite{codello2009fixed}, the effective dimension near
the UV fixed point tends to $d_{eff}=d-\eta(k\rightarrow\infty)=2$,
instead of 4, and hence in the UV region, the power of $k$ goes to
zero in Eq.(\ref{eq:general-delta}) , and $\delta_{Z}$ is expected
to behave mildly as $\log k$ instead. One could see that the dimension
$d$ and the behavior of $\delta_{Z}(k)$ are very different for small
$k$ and large $k$ region,
\begin{equation}
d_{eff}=\begin{cases}
4 & (k\rightarrow0)\\
2 & (k\rightarrow\infty)
\end{cases},\qquad\delta_{Z}\sim\begin{cases}
k^{2} & (\mathrm{for}\;\mathrm{small\;}k)\\
\log k & (\mathrm{for}\;\mathrm{large\;}k)
\end{cases}.
\end{equation}

We can see that the growth rate $\partial\delta_{Z}/\partial k$ is
always non-negative, which means $\delta_{Z}$ grows monotonously
and the coupling decreases with the increasing of $k$. However, it
does not mean the theory is asymptotically free, it is in fact ``asymptotically
safe''. The reason is transparent, although at IR $\delta_{Z}$ grows
as $\sim k^{2}$, the rate slows down to $\sim(d-2)k^{d-2}$ in the
UV region, and finally the growth rate vanishes as $d\rightarrow2$
and the coupling approaches to a finite value at the UV fixed point.

The dimension reduction mechanism from 4d to 2d in gravitational system
is crucial for its non-perturbative renormalizability and the existence
of a non-Gaussian fixed point suggested by literature \cite{weinberg1979ultraviolet,Litim:2003vp,Ambjorn:2005db,Benedetti2009Fractal,Calcagni2010Fractal,Calcagni2010Quantum,Lauscher2005Fractal,Stoica2014Metric}.
The difference between those and ours is that it is the dimension
reduction of $d$ of the base space of the NLSM but the physical spacetime
dimension $D$ of the target space, although they are related in the
semi-classical approximation. The dimension reduction of the base
space of NLSM is purely a renormalization effect which has a fractal
geometry interpretation \cite{Svozil1987Quantum,H2000Fractal}: $Z$
renormalizes the geometric measure $d^{d}x$ and makes the dimension
$d$ fractal but keeping other parameters such as $\lambda$ fixed.
Since $Z\sim k^{d-2}\sim\Delta x^{-(d-2)}$, $\Delta x$ is the cutoff
length, when $\Delta x$ tends to be infinitely small, the renormalized
geometric measure then becomes $Zd^{d}x\sim\Delta x^{-(d-2)}d^{d}x\rightarrow d^{2}x$,
so the scaling or Hausdorff dimension of the geometric measure becomes
2d. Roughly speaking, the infinity appearing in the renormalization
function can be absorbed into the re-definition of the dimension of
the base space, a 2d NLSM with bare $\lambda$ without induced Ricci
scalar correction, in the situation, is mathematically equivalent
to a 4d NLSM containing an induced Ricci scalar as a renormalization
correction to $\lambda$, depending on which quantity ($d^{d}x$ or
$\lambda$) $Z$ renormalizes in the interpretation.

Because of the mechanism, the proposed theory is very different from
conventional 2d NLSM which is effectively the UV limit of our theory,
while in the IR region the theory behaves like a 4d NSLM with Ricci
scalar correction. For this reason, the asymptotic UV critical behavior
of the theory does not explicitly depend on the dimension $d$ of
the base space. The qualitative behavior of the theory is determined
not by the explicit fundamental Lagrangian, but rather by the nature
of the basic symmetry $O(D)$ or $O(1,D-1)$, where $D$ is the only
parameter, that are imposed on the family of Lagrangian with different
$d$ that flow into one another, and by the universal nature of the
UV fixed point.

In the UV region or small scale, the classical geometric notions such
as metric or curvature of the target manifold (spacetime) may become
improper, a viable approach to probe this region may be to regard
the NLSM as an equivalent constrained linear theory Eq.(\ref{eq:NLSM-1})
without these geometric notions. In the situation, the geometric measure
of the base space can be considered being renormalized and hence the
dimension reduces to $d\rightarrow2$, the renormalization function
at UV can be conveniently evaluated 
\begin{equation}
\delta_{Z}(k)-\delta_{Z}(0)=D\int_{\Lambda\leq\left|p\right|<k}\frac{d^{2}p}{(2\pi)^{2}}\frac{i}{p^{2}},
\end{equation}
in which $\Lambda$ is an IR cutoff. We obtain the integral for the
UV cutoff $k$, 
\begin{equation}
\delta_{Z}(k)=\frac{D}{2\pi}\log\frac{k}{\Lambda}+C,\quad(\mathrm{for}\;\mathrm{large\;}k).\label{eq:delta-UV}
\end{equation}
which is a counterpart of Eq.(\ref{eq:delta-IR-1}) for large $k$
when the dimension is sufficiently reduced to $d_{eff}=2$. 

In order to determine the k-independent constant $C$, we make use
of the continuity and universality of the scaling dimension $\gamma$
function in a renormalization evolution equation. For small $k$ where
$d_{eff}=4$, by using Eq.(\ref{eq:delta-IR-1}) the $\gamma$ function
is given by
\begin{equation}
\gamma(k)=\frac{1}{2}k\frac{\partial\delta_{Z}}{\partial k}=\frac{1}{128\pi^{2}}\frac{R}{\lambda}k^{2},\quad(\mathrm{for}\;\mathrm{small}\;k),
\end{equation}
while for large $k$ where $d_{eff}\rightarrow2$, by using Eq.(\ref{eq:delta-UV})
it changes from the quadratic behavior to a universal critical scaling
dimension
\begin{equation}
\gamma_{c}=\gamma(k\rightarrow\infty)=\frac{1}{2}k\frac{\partial\delta_{Z}}{\partial k}=\frac{D}{4\pi},\quad(\mathrm{for}\;\mathrm{large}\;k).\label{eq:critical_dimension}
\end{equation}
Since $\gamma$ is a smooth function of $k$ and, in the asymptotic
UV region, it continuously approaches the universal value independent
of the cutoff $k$, they must be identical at certain large $k_{id}$,
\begin{equation}
\gamma(k_{id})=\frac{1}{128\pi^{2}}\frac{R}{\lambda}k_{id}^{2}=\frac{D}{4\pi}.\label{eq:identified_k}
\end{equation}
At such large scale $k_{id}$ where the $\gamma$ function becomes
universal and hence the scaling of the system starts to become critical,
it is safe to consider the initial renormalization condition Eq.(\ref{eq:renormalization-condition})
starts applying. Substituting the Eq.(\ref{eq:identified_k}) into
Eq.(\ref{eq:delta-IR-1}) we have
\begin{equation}
\delta_{Z}(k_{id})=\frac{D}{4\pi}+C=0,
\end{equation}
which immediately demands a universal value of $C$ relating to the
universal critical scaling $\gamma_{c}$ at UV, by using $D\equiv4$,

\begin{equation}
C=-\gamma_{c}=-\frac{1}{\pi}.\label{eq:c}
\end{equation}

Despite the above result being evaluated under the one-loop approximation,
to some extent, the general result that the constant $C$ is exactly
the minus critical scaling $-\gamma_{c}$ is reliable independent
of any approximation method. Because in the vicinity of the UV fixed
point, the system becomes nearly critical, the renormalization function
$Z(k)$ is nothing but a power behavior
\[
Z(k\rightarrow\infty)\sim\left(\frac{k^{2}}{\Lambda^{2}}\right)^{\gamma(k)+C},\quad(\mathrm{for}\;\mathrm{large}\;k).
\]
Therefore, by applying the initial renormalization condition, $Z(k\rightarrow\infty)=1$,
the exponent $\gamma(k)+C$ has to exactly vanish at UV, so it must
have $C=-\gamma_{c}$.

At this moment, we arrive at an asymptotic function $\delta_{Z}$
in the IR region, which is able to extrapolatively satisfy the initial
renormalization condition at UV,
\begin{equation}
\delta_{Z}(k)=-\frac{1}{\pi}+\frac{1}{128\pi^{2}}\frac{R}{\lambda}k^{2},\quad(\mathrm{for}\;\mathrm{small\;}k).\label{eq:delta-IR-2}
\end{equation}
For this formula, we should not incorrectly consider that $k$ must
stop at a particular point $k_{c}$ to impose the renormalization
condition ($\delta_{Z}(k_{c})=0$). We must emphasize that $k$ will
not stop at $k_{c}$, it could continuously go to infinity, since
at large $k$ the behavior of $\delta_{Z}$ changes as the effective
dimension approaches $2$ instead of 4, and $\delta_{Z}$ increases
as $\log k$ instead of $k^{2}$. We know that the NLSM near $d=2$
becomes perturbative renormalizable \cite{brezin1976renormalization,friedan1980nonlinear,Friedan1980}
which is a positive feature of our model at UV. And finally the increasing
rate slows down and stops at the UV fixed point, where the $\delta_{Z}$
vanishes as the renormalization condition imposes, leaving the finite
bare $\lambda$. In this sense, the theory truly has a non-trivial
UV fixed point, where the theory is well-defined at quantum level
with finite number of relevant bare inputs. 

On the other hand, note that the critical scaling dimension at UV
affects the IR behavior non-trivially. As $k\rightarrow0$, $Z(k)$
changes from unity at UV limit to $1-\gamma_{c}$ at IR limit. At
low energy, the semi-classical approximation requires $d=D=4$, it
is convenient to interpret that the theory has a non-trivial IR fixed
point where $\lambda$ is renormalized to a finite value at $d=4$,
\begin{equation}
\lambda_{IR}=\left(1-\frac{1}{\pi}\right)\lambda\approx0.68\lambda.
\end{equation}
Note that the $\lambda_{IR}$ is of the same order of the UV bare
value $\lambda$, so the perturbation technique is not only valid
for UV but also IR. In this sense, we consider the above results made
for small $k$ also to be reliable. As we will discuss later, the
non-trivial IR fixed point is very crucial for understanding the cosmic
observations, for instance, cosmological constant and the accelerating
expansion of universe.

\subsection{Effective Action: Emergent Cosmological Constant}

To see how the theory relates to an effective cosmological constant,
in this subsection, it is convenient to interpret the theory at low
energy when $d=4$ while $\lambda$ is renormalized. Substituting
the Eq.(\ref{eq:delta-IR-2}) into Eq.(\ref{eq:renormalized-action}),
one gets the effective action in the IR region,
\begin{equation}
S_{k}=\frac{1}{2}\int d^{4}x\left(\lambda-\frac{1}{\pi}\lambda+\frac{1}{128\pi^{2}}Rk^{2}\right)g_{\mu\nu}\partial_{a}X^{\mu}\partial_{a}X^{\nu},
\end{equation}
in which the parameter $\lambda$ is a $x$-independent constant and
hence can be taken into the integral. Remind the renormalization condition
Eq.(\ref{eq:renormalization-condition}) that near the UV fixed point
the second term in the parenthesis is nearly canceled by the asymptotic
UV form of the third term, leaving the first term $\lambda$ near
the UV fixed point an almost Einstein-Hilbert-like term. Indeed, perturbation
result \cite{codello2009fixed} shows that, the first term $\lambda$,
up to a constant factor, is $\lambda\propto R_{*}k_{UV}^{2}$, where
$R_{*}$ is a Ricci scalar curvature at a fixed point that takes a
value of classical solution, and $k_{UV}$ a UV scale, and hence the
first term $\lambda$ looks like an Einstein-Hilbert term (at a fixed
point). So formally, replacing the first term $\lambda$ with an Einstein-Hilbert-like
term, we rewrite action as 
\begin{equation}
S_{k}=\frac{1}{4}\int d^{4}x\left(\frac{R_{*}}{16\pi G}-\frac{2}{\pi}\lambda+\frac{1}{64\pi^{2}}Rk^{2}\right)g_{\mu\nu}\partial_{a}X^{\mu}\partial_{a}X^{\nu},\label{eq:zeta}
\end{equation}
where $G$ is the Newton's constant playing the role of the $k_{UV}^{-2}$,
the pre-factor $1/4$ is for convenience comparing with the semi-classical
action Eq.(\ref{eq:semi-classical-action-I}).

In this sense, in the above IR action, the first term, which is the
only leaving term at UV, is interpreted as an Einstein-Hilbert-like
term; while the second term, which is canceled at UV, is interpreted
as a cosmological constant term appearing only at IR. Although, at
first glance, the first and second term in the parenthesis are indistinguishable,
they have different meanings at UV and IR, so the decomposition of
first and second term is not arbitrary here. 

As the renormalization flow to $k\rightarrow0$, a Ricci scalar $R$
flows to a homogeneous constant curvature $R_{*}$ at the fixed point.
Thus $R_{*}$ is a k-independent constant describing a homogeneous
curvature background of the target manifold, the third k-dependent
renormalization correction term describes a curvature fluctuation
around the background. At certain renormalization scale, the first
and third terms in the parenthesis together reproduce an effective
k-dependent Einstein-Hilbert term in terms of Ricci scalar $R$. Consequently,
a locally inhomogeneous Ricci scalar $R$ flows to the homogeneous
$R_{*}$ by a flow 
\begin{equation}
R=R_{*}\left(1-\frac{1}{4\pi}Gk^{2}\right)^{-1},\quad(\mathrm{for}\;\mathrm{small}\;k),\label{eq:R*}
\end{equation}
in which we find that $R_{*}$ is in fact an IR fixed point value
of $R$. From this flow equation of Ricci scalar, we can see that
a locally inhomogeneous $R$ in the standard Einstein-Hilbert term
is reproduced from the renormalization effect. At larger and larger
spacetime scale, an inhomogeneous Ricci scalar $R$ is smoothed out
and converges to a constant background curvature $R_{*}$, or equivalently
speaking, a locally inhomogeneous Einstein-Hilbert term is reproduced
from a smooth background $R_{*}$ adding local quantum fluctuations. 

At this moment, we could substitute the effective action of reference
frame Eqs.(\ref{eq:zeta},\ref{eq:R*}) into the Eq.(\ref{eq:total-action})
and using the semi-classical approximation to treat $X_{\mu}$, then
obtain a total effective action formulated by using the internal and
physical coordinates $X_{\mu}$,
\begin{equation}
S_{eff}[\varphi,g_{\mu\nu}]=\int d^{4}X\sqrt{\mathrm{det}g}\left[\frac{1}{2}g^{\mu\nu}\frac{\delta\varphi}{\delta X^{\mu}}\frac{\delta\varphi}{\delta X^{\nu}}-V(\varphi)+\frac{R}{16\pi G}-\frac{2}{\pi}\lambda\right].\label{eq:total-effective-action}
\end{equation}
The first two terms in the bracket are the ordinary matter term, the
third term is interpreted as the Einstein-Hilbert term at certain
scale. The fourth term $(2/\pi)\lambda$ is a positive constant related
to the IR fixed point Ricci scalar being of order $R_{*}k_{UV}^{2}$,
which here is interpreted as the effective cosmological constant corresponding
to an effective energy density 
\begin{equation}
\rho_{\Lambda}=\frac{2}{\pi}\lambda.
\end{equation}

Note that the critical density is defined by $\rho_{c}=\frac{3H^{2}}{8\pi G}$,
where $H$ is the Hubble parameter. The value of near current epoch
$H_{0}=\left.H\right|_{z\sim0}$ then gives $\left.\rho_{c}\right|_{z\sim0}=\frac{3H_{0}^{2}}{8\pi G}$,
which is a density averaged by a volume with respect to rulers and
clocks at near current epoch. By using the relation between the homogeneous
Ricci scalar background and curvature radius $r$, $R_{*}\equiv D(D-1)r^{-2}=12r^{-2}$,
where the curvature radius $r$ is given by the current Hubble parameter
$r=H_{0}^{-1}$, so one finds that the critical density is just the
UV fixed point value of $\lambda$, 
\begin{equation}
\left.\rho_{c}\right|_{z\sim0}=\lambda=\frac{R_{*}}{32\pi G}.
\end{equation}
Then we have $\rho_{\Lambda}\doteq(2/\pi)\left.\rho_{c}\right|_{z\sim0}$,
where ``$\doteq$'' stands for the neglecting of the renormalization
correction. It is an interesting result that the dark energy is equal
to $2/\pi$ times the ``current'' critical density,

\begin{equation}
\left.\Omega_{\Lambda}\right|_{z\sim0}=\frac{2}{\pi}\approx0.64,
\end{equation}
which agrees with the observations well. The word ``current'' means
that the density is averaged by the volume relative to the scales
of rulers and clocks near $z\sim0$. Certainly, $\rho_{\Lambda}\doteq(2/\pi)\lambda$
and $(2/\pi)\left.\rho_{c}\right|_{z\sim0}$ are just equal in values,
they are not really identical. $\rho_{\Lambda}\doteq(2/\pi)\lambda\propto R_{*}$
is a scalar, it gives rise to a full stress tensor $T_{\mu\nu}=\rho_{\Lambda}g_{\mu\nu}$,
for this reason, its equation of state is exactly $w=-1$, in contrast,
$\rho_{c}$ is just the $T_{00}$ component of a stress tensor, so
they behave differently under spacetime coordinates transformation.
It is easy to see that $\rho_{\Lambda}\doteq(2/\pi)\lambda\propto R_{*}$
is invariant with respect to the physical clock time, in this sense
it is a constant, but $(2/\pi)\rho_{c}$ varies with the redshift,
they are just equal in values near $z\sim0$.

In the framework only the concepts defined in a relative way, instead
of absolute, are observable. A physical observable is redshift $z$,
while the absolute global age of the universe is essentially unobservable
in our theory. In this sense, discussions about the universe evolution
should not be based on the global age but the redshift. In other words,
the quantities being functions of the global age, e.g. $H(t)$, $\rho_{i}(t)$,
$\Omega_{i}(t)$ should be replaced by some more physical ones being
functions of the redshift: $H(z)$, $\rho_{i}(z)$, $\Omega_{i}(z)$.
It is no problem that the near current epoch $z\sim0$ always exist
in every epoch of the expansion history of the universe, leading to
the consequence that an observer at his/her near ``current'' epoch
``always'' find $\left.\Omega_{\Lambda}\right|_{z\sim0}\approx0.64$
no matter what is the absolute epoch he/she lives in. In this sense,
$\rho_{\Lambda}$ is ``always'' comparable with $\left.\rho_{c}\right|_{z\sim0}$
seen by observers at his/her epoch. The word ``always'' is about
the relational redshift instead of the absolute history or age, meaning
that the value of $\Omega_{\Lambda}$ is fixed with respect to the
observers at $z\sim0$, it does not mean $\rho_{\Lambda}$ and $\rho_{c}$
behave in the same way in the expansion history of the universe. As
we know, they behave differently about the redshift, $\rho_{c}(z)$
increases with $z$, while $\rho_{\Lambda}$ does not change, resulting
to that $\Omega_{\Lambda}(z)=\rho_{\Lambda}/\rho_{c}(z)$ decreases
as redshift increases. The framework gives a non-dynamical explanation
to the coincidence problem, namely why the dark energy is comparable
with the critical density now? The key to understanding the coincidence
is to use the relational observable (such as redshift) independent
of any absolute scale of the universe, avoiding using the quantity
such as the global age of the universe defined by an absolute observer
looking from outside. We will also come back to the coincidence problem
in discussing the distance-redshift relation in the next subsection.

\subsection{Physical Interpretation: Spacetime Uncertainties at Cosmic Distance}

We have seen that the renormalization function $\delta_{Z}$ or $Z$
is crucial in understanding the cosmological constant. To see how
it relates to cosmic observational effect, in this subsection, it
is convenient to interpret the effective theory at $d=4$ and $\lambda$
fixed, while equivalently, spacetime coordinates $X_{\mu}$ are isotropically
renormalized by the renormalization function.

The previous renormalization results can also be understood by the
mean field or background field method. In such language, providing
that $\Delta X_{\mu}(x)$ is a displacement between $X_{\mu}(x)$
and a given point $X_{\mu}(0)$, i.e. $\Delta X_{\mu}(x)=X_{\mu}(x)-X_{\mu}(0)$.
The renormalized quantity $\Delta X_{\mu}^{re}$ can be expanded around
the mean field value $\langle\Delta X_{\mu}\rangle$ by a quantum
fluctuation $\delta X_{\mu}$, 
\begin{equation}
\Delta X_{\mu}^{re}=\langle\Delta X_{\mu}\rangle+\delta X_{\mu}.
\end{equation}
By using this relation and Eqs.(\ref{eq:definition-Z},\ref{eq:renormalized-action}),
considering that the vacuum expectation value of the fluctuation vanishes
i.e. $\langle\delta X_{\mu}\rangle=0$, we find that $\delta_{Z}(k)$
is nothing but a measure of a ratio of each variance to the mean-squared
of its spacetime coordinate difference, 
\begin{equation}
\delta_{Z}(k)=Z(k)-1=-\frac{1}{2}\frac{\langle\delta X_{\mu}^{2}\rangle}{\langle\Delta X_{\mu}\rangle^{2}}.
\end{equation}
Recall the result Eq.(\ref{eq:delta-IR-2}), we have
\begin{equation}
\frac{\langle\delta X_{\mu}^{2}\rangle}{\langle\Delta X_{\mu}\rangle^{2}}=-2\delta_{Z}=\frac{2}{\pi}\left(1-\frac{1}{4}Gk^{2}\right),\quad(\mathrm{for}\;\mathrm{small}\;k).
\end{equation}

This formula indicates an inescapable and universal quantum limitation
to the spacetime accuracy at IR. One can not have rulers and clocks
precisely measured and synchronized across spacetime. In the IR limit,
or equivalently, at cosmic scale, the ratio of the variance $\langle\delta X_{\mu}^{2}\rangle$
to the mean-squared of spacetime distance $\langle\Delta X_{\mu}\rangle^{2}$
is universal, 
\begin{equation}
\lim_{k\rightarrow0}\frac{\langle\delta X_{\mu}^{2}\rangle}{\langle\Delta X_{\mu}\rangle^{2}}=\frac{2}{\pi}.\label{eq:universal-ratio}
\end{equation}
The universality of the ratio is closely related to the universality
of the critical dimension $\gamma_{c}$ in Eq.(\ref{eq:c}). In the
next subsection, we will see that the universality of the ratio is
also a natural consequence of the generalized equivalence principle.

The variance $\langle\delta X_{\mu}^{2}\rangle$ in Eq.(\ref{eq:universal-ratio})
is valid at extreme IR or cosmic distance $\langle\Delta X_{\mu}\rangle^{2}$,
which can be seen as an inherent cosmic variance of a quantum measurement
at cosmic distance. It generalizes the result of refs.\cite{2014NuPhB.884..344L,2015JHEP...06..063L}
by putting the space and time on an equal footing. In those papers,
only the uncertainty of a physical clock time field $X_{0}$ is considered,
nevertheless, the spatial coordinates are just treated as parameter
background for the sake of simplicity. The difference between them
is as follows. For the result in refs.\cite{2014NuPhB.884..344L,2015JHEP...06..063L},
the clock time variance grows linearly with the spatial distance and
the proportional coefficient is the inverse of the spatial volume
cutoff. However, in this paper, the clock variance instead grows quadratically
instead of linearly. Since the space and time are considered symmetric
and isotropic, the clock variance can be interpreted as being proportional
to the squared spatial distance. The reason for the different power
dependence is due to the fact that, in the standard quantum mechanics,
it is of first order in the derivative with respect to an evolution
parameter, for instance the derivative with respect to the Schrodinger's
time or the renormalization scale $\log k$, but in a theory that
space and time are put on an equal footing, for instance in a relativistic
theory, the orders in the derivatives with respect to spatial and
temporal distance are the same, for bosonic degrees of freedom, they
are both of second order.

For the reason that time or frequency can be conveniently measured
by redshifts of distant spectral lines in cosmic observations, they
could be used as idea clocks distant from us to test the effect at
cosmic distance. The clock time uncertainty can also be interpreted
in terms of a redshift uncertainty or broadening. Although many physics
affect the redshift broadening such as thermal fluctuation, a prediction
may be testable: the ratio of the inherent quantum variance of redshift
to its mean-squared equals 
\begin{equation}
\lim_{k\rightarrow0}\frac{\langle\delta z^{2}\rangle}{\langle z\rangle^{2}}=\frac{2}{\pi},\label{eq:universal-ratio-redshift}
\end{equation}
which is derived from Eq.(\ref{eq:universal-ratio}) by using $\left\langle \delta z^{2}\right\rangle =\left\langle \delta X_{\mu}^{2}\right\rangle /\left\langle X_{\mu}(0)\right\rangle ^{2}$
and $\left\langle z\right\rangle =\left\langle X_{\mu}(x)-X_{\mu}(0)\right\rangle /\left\langle X_{\mu}(0)\right\rangle $
if we consider $\langle X_{\mu}(x)\rangle$ measures certain characteristic
wavelength or period at the given point $x$. The result is also deduced
in ref.\cite{2015JHEP...06..063L}. The $\langle\delta z^{2}\rangle$
does not change the mean value $\langle z\rangle$, but shifts the
mean-squared 
\begin{equation}
\langle z^{2}\rangle=\langle z\rangle^{2}+\langle\delta z^{2}\rangle\overset{k\rightarrow0}{=}\left(1+\frac{2}{\pi}\right)\langle z\rangle^{2}.
\end{equation}
Therefore, the distance-redshift relation $D(z)$ is modified at order
$\mathcal{O}(z^{2})$ by this effect \cite{2015JHEP...06..063L}.
The first order $\mathcal{O}(z)$ term of $D(z)$ relates to the expansion
rate of the universe, and the second order $\mathcal{O}(z^{2})$ term
relates to an accelerating or decelerating of the expansion. More
precisely, the universal ratio Eq.(\ref{eq:universal-ratio-redshift})
contributes an additional deceleration parameter $q_{0}=-2/\pi$,
which is redshift independent and uniform, to the distance-redshift
relation besides other components of the universe. It makes a departure
to the Hubble's law which is more significant at high redshift. Expanding
the expectation value of the luminosity distance $\langle D(z)\rangle$
in powers of redshift to the second order \cite{Weinberg2008}, we
have 
\begin{equation}
\langle D(z)\rangle=\frac{1}{H_{0}}\left[\langle z\rangle+\frac{1}{2}\left(1+\frac{2}{\pi}+...\right)\langle z\rangle^{2}+\mathcal{O}(\left\langle z\right\rangle ^{3})\right],\label{eq:distance-redshift-relation}
\end{equation}
where the additional positive constant $2/\pi$ plays the role of
the percentage of dark energy $\Omega_{\Lambda}$, the $...$ in the
parenthesis represents the deceleration parameter coming from other
components of the universe such as ordinary matter given by $-\Omega_{m}(1+z)^{3}$.
$H_{0}$ is the Hubble constant measured at $z\sim0$. The formula
does not involve any absolute age of the universe, so for any observer,
it is always valid no matter when the observer lives in the expansion
history of the universe, the universe is always seen to become accelerating
at near current epoch $z\approx0.3$ at which the dark energy and
the matter become comparable, i.e. $q_{0}=\Omega_{m}(1+z)^{3}-\Omega_{\Lambda}=0$.
This fact also demonstrates the coincidence. The $\left\langle \delta z^{2}\right\rangle $
correction to the distance-redshift relation makes it anomalous at
high redshift, which is observed from high redshift supernovas being
the first indication of the accelerating expansion of the universe
\cite{Riess:1998cb,Perlmutter:1998np}.

\subsection{Generalization of the Equivalence Principle}

In previous sections, we have shown that a correct value of cosmological
constant and an effective Einstein-Hilbert term emerge in the IR region
of the quantum dynamics of spacetime reference frame. From the dynamics
of quantum spacetime reference frame to the concept of gravity, a
further assumption is required: the equivalence principle. The principle
gives a spacetime interpretation of gravity in classical general relativity,
which is well established and tested in classical physics. However,
the principle puzzles physicist when the quantum effects are seriously
taken into account. Since the zero-point quantum fluctuation seems
real (e.g. contributing to the Lamb shift), why these large amounts
of energies of vacuum do not gravitate as the equivalence principle
asserts, is the main puzzle of the cosmological constant problem.

In fact the Lamb shift gravitates normally \cite{2006hep.th....3249P,2009PhLB..679..433M},
and there is no hint to assume that the energies coming from classical
and quantum contributions produce different gravitational effects.
In fact, only when the equivalence principle is valid, the accelerating
expansion of the universe revealed by Eq.(\ref{eq:distance-redshift-relation})
is equivalent to the existence of a ``dark energy'' or a positive
non-vanishing cosmological constant revealed by Eq.(\ref{eq:total-effective-action}),
so it seems like an indication that in our framework the equivalence
principle could exactly hold. Furthermore, an elegant and economic
assumption is also to maintain the spirit of relativity and hence
the equivalence principle, so that gravity is just simply a relative
phenomena, and there is nothing else more than that even at the quantum
level. In this sense, the equivalence principle must be generalized
to the quantum level to resolve the cosmological constant problem.

In this paper, we argue that the equivalence principle is also valid
at the quantum level as it applies to the classical level. As the
generalized equivalence principle claims, all kinds of energies including
the quantum fluctuations gravitate. The generalized equivalence principle
implies that all kinds of apparent curving of spacetime including
those coming from quantum fluctuations or quantum uncertainties of
spacetime are equivalent to gravitation, more precisely, the quantum
uncertainties of spacetime deduced in this paper are equivalent to
the accelerating expansion of the universe. Not only one can not distinguish
gravitation from acceleration according to the classical equivalence
principle, but is also unable to distinguish gravitation from quantum
spacetime intrinsic uncertainty/fluctuation. Therefore, the Ricci
scalar in the effective action Eq.(\ref{eq:total-effective-action})
can precisely be interpreted as gravity. The accuracy of the generalized
equivalence principle can be demonstrated by the universality of the
ratio Eq.(\ref{eq:universal-ratio-redshift}) which is independent
of the energies of the spectral lines. In other words, all spectral
lines taking different energies uniformly ``free-fall''. It is not
merely a particular property of the spectral lines, it is a universal
property of the spacetime itself. So, a universal accelerating expansion
inevitably appears. The uncertainty/fluctuation of spacetime Eq.(\ref{eq:universal-ratio})
or redshift Eq.(\ref{eq:universal-ratio-redshift}) on the one hand
can be interpreted as that the objects (spectral lines) are uniformly
accelerating, or equivalently on the other hand, the spacetime is
curved by the quantum fluctuation energy density $(2/\pi)\lambda$
in Eq.(\ref{eq:total-effective-action}) which is seen as a repulsive
gravitational force.

To understand further how the generalized equivalence principle resolves
the notorious problem of zero-point vacuum energies $\sum_{k}\frac{1}{2}\omega_{k}$
predicted in textbook quantum fields theory, we note that such zero-point
vacuum energies are not involved in our effective theory Eq.(\ref{eq:total-effective-action}).
The effective vacuum energies' density or the cosmological constant
as a source of gravity in Eq.(\ref{eq:total-effective-action}) comes
from the two-point function $\langle\delta X_{\mu}\delta X_{\nu}\rangle\neq0$,
while the vacuum expectation value vanishes, $\langle\delta X_{\mu}\rangle=0$.
In other words, the energies of quantum fluctuations of spacetime
relating to two states are the leading contribution to the vacuum
energies and gravitational effects \cite{padmanabhan1987response,2005CQGra..22L.107P},
which do obey the generalized equivalence principle. However, the
notorious zero-point vacuum energies relating to one state have no
gravitational effects. Since quantum fluctuation of a physical reference
frame is inevitable, fundamentally speaking, the absolute rest frame
can not be precisely realized at the quantum level. For this reason,
the zero-point vacuum energies which make sense only with respect
to a perfect classical absolute rest frame in textbook quantum mechanics
are completely unphysical and unobservable in any laboratory experiment,
including the Casimir effect \cite{PhysRevD.72.021301}. This property
resolves the first part of the cosmological constant problem, namely,
why the zero-point vacuum energies do not gravitate. The physical
energies which gravitate normally are those make sense with respect
to the quantum spacetime reference frame that is also subject to quantum
fluctuation. As an application, the dark energy from the vacuum fluctuation
$\langle\delta X_{\mu}\delta X_{\nu}\rangle$, which relative to each
other, gravitate normally as the generalized equivalence principle
asserts. This property solves the second part of the cosmological
constant problem, namely, why the cosmological constant is so small.

\section{Summary and Conclusions}

To solve the cosmological constant problem, the quantum dynamics and
the effects of a quantum clock can not be neglected. In this paper,
acting on the spirit of treating space and time on an equal footing,
we generalize the quantum clock to the quantum spacetime reference
frame, via a physical realization of a reference system by quantum
rulers and clocks. It is in this sense we have a ``quantum spacetime''
obeying both quantum mechanics and general relativity.

In order to accommodate quantum mechanics to general relativity, the
textbook quantum mechanics must be generalized. General relativity
is general covariant or observer independent, a physical quantum state
satisfying this property is in general an entangled state entangling
a to-be-studied quantum physical system with a quantum measuring device.
It is necessary to have a relational interpretation to the entangled
state, since the to-be-studied system makes sense only relative to
the quantum measuring device. Entangled state solves the diffeomorphism
constraints and the Wheeler-DeWitt equation, which plays a more fundamental
role than the textbook Schrodinger equation. Since clock time is inescapably
subject to quantum fluctuation, the Schrodinger equation using the
parameter time is just an approximation. The cosmological constant
problem is an indication of going beyond the Schrodinger equation
where the quantum fluctuation of time is inescapable and must not
be ignored.

Omitting the internal degrees of freedom of the physical rulers and
clocks, such as their spins, considering only their metric properties,
the spacetime reference frame is described by the bosonic NLSM. In
a semi-classical treatment of the spacetime reference frame, we recover
the existing theories: quantum fields theories in fixed curved spacetime.
In a complete quantum treatment, we studied its normalization behavior
under approximations. The theory has a non-trivial UV fixed point,
namely it is asymptotically safe, and hence the notion of spacetime
still makes sense even at UV. We get three surprising results from
the theory. The first, and most remarkable, result is that a cosmological
constant appears which naturally gives not only a correct order but
also a percentage $\Omega_{\Lambda}=2/\pi\approx0.64$ very close
to current observations. The second result is that the quantum dynamics
of the quantum spacetime automatically contains an effective Einstein-Hilbert
action, and hence automatically incorporates a theory of gravity under
the assumption of the validity of equivalence principle at the quantum
level. The third result says that the spacetime are inescapably subject
to quantum uncertainties, one can not have rulers and clocks perfectly
measured and synchronized across spacetime. The ratio of the variance
to the mean-squared of the spacetime distance tends to a universal
constant $\langle\delta X_{\mu}^{2}\rangle/\langle\Delta X_{\mu}\rangle^{2}=2/\pi$
in the extreme IR region of theory. We also argue that this effect
is testable by observing a linear dependence between the inherent
quantum variance and mean-squared of redshifts from distant spectral
lines. The proportionality is $\mathcal{O}(1)$ and expected to be
identical to the percentage of the dark energy $\Omega_{\Lambda}$.
These results strongly support the argument that the equivalence principle
still holds at the quantum level. It is in this sense the theory is
possibly a good starting point of a ``quantum theory of gravity''. 

Undeniably, there are limitations in our argument, we discuss some
of them here. First, although there are several indications (e.g.
anomalous dimension, geometric measure renormalization of the base
space, and the fractal behavior), a dimension reduction, from a 4d
at IR to a 2d NLSM at UV that relates to the existence of the UV fixed
point and some of our relevant calculations, is just formally proved,
a strict proof is not given. Second, although the semi-classical theory
of matter fields in the dynamical spacetime is given, how the matter
fields renormalize when the spacetime coordinates are dynamical is
not discussed. Roughly speaking, if we rescale the usual parameter
spacetime coordinates, the dynamical coordinates rescales in the same
way at the leading order which can be seen in the semi-classical approximation.
The quantum effect of the dynamical coordinates smears the coordinates
and hence gives an additional correction to a standard renormalizing
quantity at the order $O(\delta x^{2})$. The process has not been
worked out in detail, the question is beyond the scope of the paper
and leaves for future studies: to what extent the standard renormalization
procedure of matter fields require justification in the framework?
Third, the existence of a non-trivial UV fixed point of Eq.(\ref{eq:NLSM-UV})
when $d=4$ is believed in literature only at perturbative level,
a non-perturbative proof is still lack. Whether the theory really
renormalizable needs strict proof. Fourth, our discussion to the renormalization
of NLSM relevant to the cosmological constant is analogous to a normalized
Ricci-type flow \cite{chow2004ricci} with positive isotropic initial
condition which is free from singularity, a more general initial condition
to the Ricc-type flow equation, when local singularities may develop,
is beyond our discussion. Whether the RG flow solution of the theory
always exist in any circumstance, is also a not-fully-understood question
relating to whether the theory is well-defined at the quantum level.
\begin{acknowledgments}
This work was supported in part by the National Science Foundation
of China (NSFC) under Grant No.11205149, and Science Research Foundation
of Jiangsu University under Grant No.15JDG153.

\bibliographystyle{plain}

\nocite{*}

\begin{thebibliography}{45}
\expandafter\ifx\csname natexlab\endcsname\relax\def\natexlab#1{#1}\fi
\expandafter\ifx\csname bibnamefont\endcsname\relax
  \def\bibnamefont#1{#1}\fi
\expandafter\ifx\csname bibfnamefont\endcsname\relax
  \def\bibfnamefont#1{#1}\fi
\expandafter\ifx\csname citenamefont\endcsname\relax
  \def\citenamefont#1{#1}\fi
\expandafter\ifx\csname url\endcsname\relax
  \def\url#1{\texttt{#1}}\fi
\expandafter\ifx\csname urlprefix\endcsname\relax\def\urlprefix{URL }\fi
\providecommand{\bibinfo}[2]{#2}
\providecommand{\eprint}[2][]{\url{#2}}

\bibitem[{\citenamefont{Weinberg}(1989)}]{RevModPhys.61.1}
\bibinfo{author}{\bibfnamefont{S.}~\bibnamefont{Weinberg}},
  \bibinfo{journal}{Rev. Mod. Phys.} \textbf{\bibinfo{volume}{61}},
  \bibinfo{pages}{1} (\bibinfo{year}{1989}).

\bibitem[{\citenamefont{{Bousso}}(2008)}]{2008GReGr..40..607B}
\bibinfo{author}{\bibfnamefont{R.}~\bibnamefont{{Bousso}}},
  \bibinfo{journal}{General Relativity and Gravitation}
  \textbf{\bibinfo{volume}{40}}, \bibinfo{pages}{607} (\bibinfo{year}{2008}),
  \eprint{0708.4231}.

\bibitem[{\citenamefont{{Martin}}(2012)}]{2012CRPhy..13..566M}
\bibinfo{author}{\bibfnamefont{J.}~\bibnamefont{{Martin}}},
  \bibinfo{journal}{Comptes Rendus Physique} \textbf{\bibinfo{volume}{13}},
  \bibinfo{pages}{566} (\bibinfo{year}{2012}), \eprint{1205.3365}.

\bibitem[{\citenamefont{{Luo}}(2014)}]{2014NuPhB.884..344L}
\bibinfo{author}{\bibfnamefont{M.~J.} \bibnamefont{{Luo}}},
  \bibinfo{journal}{Nuclear Physics B} \textbf{\bibinfo{volume}{884}},
  \bibinfo{pages}{344} (\bibinfo{year}{2014}), \eprint{1312.2759}.

\bibitem[{\citenamefont{{Luo}}(2015)}]{2015JHEP...06..063L}
\bibinfo{author}{\bibfnamefont{M.~J.} \bibnamefont{{Luo}}},
  \bibinfo{journal}{Journal of High Energy Physics}
  \textbf{\bibinfo{volume}{6}}, \bibinfo{pages}{63} (\bibinfo{year}{2015}),
  \eprint{1401.2488}.

\bibitem[{\citenamefont{{Planck Collaboration}
  et~al.}(2013)\citenamefont{{Planck Collaboration}, {Ade}, {Aghanim},
  {Armitage-Caplan}, {Arnaud}, {Ashdown}, {Atrio-Barandela}, {Aumont},
  {Baccigalupi}, {Banday} et~al.}}]{2013arXiv1303.5062P}
\bibinfo{author}{\bibnamefont{{Planck Collaboration}}},
  \bibinfo{author}{\bibfnamefont{P.~A.~R.} \bibnamefont{{Ade}}},
  \bibinfo{author}{\bibfnamefont{N.}~\bibnamefont{{Aghanim}}},
  \bibinfo{author}{\bibfnamefont{C.}~\bibnamefont{{Armitage-Caplan}}},
  \bibinfo{author}{\bibfnamefont{M.}~\bibnamefont{{Arnaud}}},
  \bibinfo{author}{\bibfnamefont{M.}~\bibnamefont{{Ashdown}}},
  \bibinfo{author}{\bibfnamefont{F.}~\bibnamefont{{Atrio-Barandela}}},
  \bibinfo{author}{\bibfnamefont{J.}~\bibnamefont{{Aumont}}},
  \bibinfo{author}{\bibfnamefont{C.}~\bibnamefont{{Baccigalupi}}},
  \bibinfo{author}{\bibfnamefont{A.~J.} \bibnamefont{{Banday}}},
  \bibnamefont{et~al.}, \bibinfo{journal}{ArXiv e-prints}
  (\bibinfo{year}{2013}), \eprint{1303.5062}.

\bibitem[{\citenamefont{{Aharonov} and {Kaufherr}}(1984)}]{1984PhRvD..30..368A}
\bibinfo{author}{\bibfnamefont{Y.}~\bibnamefont{{Aharonov}}} \bibnamefont{and}
  \bibinfo{author}{\bibfnamefont{T.}~\bibnamefont{{Kaufherr}}},
  \bibinfo{journal}{\prd} \textbf{\bibinfo{volume}{30}}, \bibinfo{pages}{368}
  (\bibinfo{year}{1984}).

\bibitem[{\citenamefont{Rovelli}(1991)}]{rovelli1991quantum}
\bibinfo{author}{\bibfnamefont{C.}~\bibnamefont{Rovelli}},
  \bibinfo{journal}{Classical and Quantum Gravity}
  \textbf{\bibinfo{volume}{8}}, \bibinfo{pages}{317} (\bibinfo{year}{1991}).

\bibitem[{\citenamefont{Dickson}(2004)}]{dickson2004view}
\bibinfo{author}{\bibfnamefont{M.}~\bibnamefont{Dickson}},
  \bibinfo{journal}{Studies in History and Philosophy of Science Part B:
  Studies in History and Philosophy of Modern Physics}
  \textbf{\bibinfo{volume}{35}}, \bibinfo{pages}{195} (\bibinfo{year}{2004}).

\bibitem[{\citenamefont{Angelo et~al.}(2011)\citenamefont{Angelo, Brunner,
  Popescu, Short, and Skrzypczyk}}]{angelo2011physics}
\bibinfo{author}{\bibfnamefont{R.~M.} \bibnamefont{Angelo}},
  \bibinfo{author}{\bibfnamefont{N.}~\bibnamefont{Brunner}},
  \bibinfo{author}{\bibfnamefont{S.}~\bibnamefont{Popescu}},
  \bibinfo{author}{\bibfnamefont{A.~J.} \bibnamefont{Short}}, \bibnamefont{and}
  \bibinfo{author}{\bibfnamefont{P.}~\bibnamefont{Skrzypczyk}},
  \bibinfo{journal}{Journal of Physics A: Mathematical and Theoretical}
  \textbf{\bibinfo{volume}{44}}, \bibinfo{pages}{145304}
  (\bibinfo{year}{2011}).

\bibitem[{\citenamefont{Page and Wootters}(1983)}]{PhysRevD.27.2885}
\bibinfo{author}{\bibfnamefont{D.~N.} \bibnamefont{Page}} \bibnamefont{and}
  \bibinfo{author}{\bibfnamefont{W.~K.} \bibnamefont{Wootters}},
  \bibinfo{journal}{Phys. Rev. D} \textbf{\bibinfo{volume}{27}},
  \bibinfo{pages}{2885} (\bibinfo{year}{1983}).

\bibitem[{\citenamefont{{Moreva} et~al.}(2014)\citenamefont{{Moreva}, {Brida},
  {Gramegna}, {Giovannetti}, {Maccone}, and {Genovese}}}]{2014PhRvA..89e2122M}
\bibinfo{author}{\bibfnamefont{E.}~\bibnamefont{{Moreva}}},
  \bibinfo{author}{\bibfnamefont{G.}~\bibnamefont{{Brida}}},
  \bibinfo{author}{\bibfnamefont{M.}~\bibnamefont{{Gramegna}}},
  \bibinfo{author}{\bibfnamefont{V.}~\bibnamefont{{Giovannetti}}},
  \bibinfo{author}{\bibfnamefont{L.}~\bibnamefont{{Maccone}}},
  \bibnamefont{and}
  \bibinfo{author}{\bibfnamefont{M.}~\bibnamefont{{Genovese}}},
  \bibinfo{journal}{\pra} \textbf{\bibinfo{volume}{89}}, \bibinfo{eid}{052122}
  (\bibinfo{year}{2014}), \eprint{1310.4691}.

\bibitem[{\citenamefont{{Rovelli}}(1996)}]{1996IJTP...35.1637R}
\bibinfo{author}{\bibfnamefont{C.}~\bibnamefont{{Rovelli}}},
  \bibinfo{journal}{International Journal of Theoretical Physics}
  \textbf{\bibinfo{volume}{35}}, \bibinfo{pages}{1637} (\bibinfo{year}{1996}),
  \eprint{quant-ph/9609002}.

\bibitem[{\citenamefont{Rovelli}(2004)}]{rovelli2004quantum}
\bibinfo{author}{\bibfnamefont{C.}~\bibnamefont{Rovelli}},
  \emph{\bibinfo{title}{Quantum gravity}} (\bibinfo{publisher}{Cambridge
  University Press}, \bibinfo{year}{2004}).

\bibitem[{\citenamefont{Phillips and Hu}(2000)}]{phillips2000vacuum}
\bibinfo{author}{\bibfnamefont{N.~G.} \bibnamefont{Phillips}} \bibnamefont{and}
  \bibinfo{author}{\bibfnamefont{B.}~\bibnamefont{Hu}},
  \bibinfo{journal}{Physical Review D} \textbf{\bibinfo{volume}{62}},
  \bibinfo{pages}{084017} (\bibinfo{year}{2000}).

\bibitem[{\citenamefont{Gell-Mann and L{\'e}vy}(1960)}]{gell1960axial}
\bibinfo{author}{\bibfnamefont{M.}~\bibnamefont{Gell-Mann}} \bibnamefont{and}
  \bibinfo{author}{\bibfnamefont{M.}~\bibnamefont{L{\'e}vy}},
  \bibinfo{journal}{Il Nuovo Cimento} \textbf{\bibinfo{volume}{16}},
  \bibinfo{pages}{705} (\bibinfo{year}{1960}).

\bibitem[{\citenamefont{Ketov}(2000)}]{ketov2000quantum}
\bibinfo{author}{\bibfnamefont{S.~V.} \bibnamefont{Ketov}},
  \emph{\bibinfo{title}{Quantum non-linear sigma-models: from quantum field
  theory to supersymmetry, conformal field theory, black holes and strings}}
  (\bibinfo{publisher}{Springer Science \& Business Media},
  \bibinfo{year}{2000}).

\bibitem[{\citenamefont{Zinn-Justin}(2002)}]{zinn2002quantum}
\bibinfo{author}{\bibfnamefont{J.}~\bibnamefont{Zinn-Justin}},
  \emph{\bibinfo{title}{Quantum field theory and critical phenomena}}
  (\bibinfo{publisher}{Oxford University Press, Oxford}, \bibinfo{year}{2002}).

\bibitem[{\citenamefont{Omero and Percacci}(1980)}]{Omero1980Generalized}
\bibinfo{author}{\bibfnamefont{C.}~\bibnamefont{Omero}} \bibnamefont{and}
  \bibinfo{author}{\bibfnamefont{R.}~\bibnamefont{Percacci}},
  \bibinfo{journal}{Nuclear Physics B} \textbf{\bibinfo{volume}{165}},
  \bibinfo{pages}{351} (\bibinfo{year}{1980}).

\bibitem[{\citenamefont{Gell-Mann and Zwiebach}(1985)}]{Gell1985Dimensional}
\bibinfo{author}{\bibfnamefont{M.}~\bibnamefont{Gell-Mann}} \bibnamefont{and}
  \bibinfo{author}{\bibfnamefont{B.}~\bibnamefont{Zwiebach}},
  \bibinfo{journal}{Nuclear Physics B} \textbf{\bibinfo{volume}{260}},
  \bibinfo{pages}{569} (\bibinfo{year}{1985}).

\bibitem[{\citenamefont{{Giddings} et~al.}(2006)\citenamefont{{Giddings},
  {Marolf}, and {Hartle}}}]{2006PhRvD..74f4018G}
\bibinfo{author}{\bibfnamefont{S.~B.} \bibnamefont{{Giddings}}},
  \bibinfo{author}{\bibfnamefont{D.}~\bibnamefont{{Marolf}}}, \bibnamefont{and}
  \bibinfo{author}{\bibfnamefont{J.~B.} \bibnamefont{{Hartle}}},
  \bibinfo{journal}{\prd} \textbf{\bibinfo{volume}{74}}, \bibinfo{eid}{064018}
  (\bibinfo{year}{2006}), \eprint{hep-th/0512200}.

\bibitem[{\citenamefont{Percacci and Roberto}(2014)}]{Percacci2014Geometry}
\bibinfo{author}{\bibnamefont{Percacci}} \bibnamefont{and}
  \bibinfo{author}{\bibnamefont{Roberto}}, \emph{\bibinfo{title}{Geometry of
  nonlinear field theories}} (\bibinfo{publisher}{World Scientific},
  \bibinfo{year}{2014}).

\bibitem[{\citenamefont{Codello and Percacci}(2009)}]{codello2009fixed}
\bibinfo{author}{\bibfnamefont{A.}~\bibnamefont{Codello}} \bibnamefont{and}
  \bibinfo{author}{\bibfnamefont{R.}~\bibnamefont{Percacci}},
  \bibinfo{journal}{Physics Letters B} \textbf{\bibinfo{volume}{672}},
  \bibinfo{pages}{280} (\bibinfo{year}{2009}).

\bibitem[{\citenamefont{Weinberg}(1979)}]{weinberg1979ultraviolet}
\bibinfo{author}{\bibfnamefont{S.}~\bibnamefont{Weinberg}}, in
  \emph{\bibinfo{booktitle}{S. W. Hawking and W. Israel (Eds), General
  relativity: An Einstein centenary survey}} (\bibinfo{publisher}{Cambridge
  University Press}, \bibinfo{year}{1979}).

\bibitem[{\citenamefont{Litim}(2004)}]{Litim:2003vp}
\bibinfo{author}{\bibfnamefont{D.~F.} \bibnamefont{Litim}},
  \bibinfo{journal}{Phys. Rev. Lett.} \textbf{\bibinfo{volume}{92}},
  \bibinfo{pages}{201301} (\bibinfo{year}{2004}), \eprint{hep-th/0312114}.

\bibitem[{\citenamefont{Ambjorn et~al.}(2005)\citenamefont{Ambjorn, Jurkiewicz,
  and Loll}}]{Ambjorn:2005db}
\bibinfo{author}{\bibfnamefont{J.}~\bibnamefont{Ambjorn}},
  \bibinfo{author}{\bibfnamefont{J.}~\bibnamefont{Jurkiewicz}},
  \bibnamefont{and} \bibinfo{author}{\bibfnamefont{R.}~\bibnamefont{Loll}},
  \bibinfo{journal}{Phys. Rev. Lett.} \textbf{\bibinfo{volume}{95}},
  \bibinfo{pages}{171301} (\bibinfo{year}{2005}), \eprint{hep-th/0505113}.

\bibitem[{\citenamefont{Benedetti}(2009)}]{Benedetti2009Fractal}
\bibinfo{author}{\bibfnamefont{D.}~\bibnamefont{Benedetti}},
  \bibinfo{journal}{Physical Review Letters} \textbf{\bibinfo{volume}{102}},
  \bibinfo{pages}{111303} (\bibinfo{year}{2009}).

\bibitem[{\citenamefont{Calcagni}(2010{\natexlab{a}})}]{Calcagni2010Fractal}
\bibinfo{author}{\bibfnamefont{G.}~\bibnamefont{Calcagni}},
  \bibinfo{journal}{Physical Review Letters} \textbf{\bibinfo{volume}{104}},
  \bibinfo{pages}{251301} (\bibinfo{year}{2010}{\natexlab{a}}).

\bibitem[{\citenamefont{Calcagni}(2010{\natexlab{b}})}]{Calcagni2010Quantum}
\bibinfo{author}{\bibfnamefont{G.}~\bibnamefont{Calcagni}},
  \bibinfo{journal}{Journal of High Energy Physics}
  \textbf{\bibinfo{volume}{2010}}, \bibinfo{pages}{1}
  (\bibinfo{year}{2010}{\natexlab{b}}).

\bibitem[{\citenamefont{{Lauscher} and {Reuter}}(2005)}]{Lauscher2005Fractal}
\bibinfo{author}{\bibfnamefont{O.}~\bibnamefont{{Lauscher}}} \bibnamefont{and}
  \bibinfo{author}{\bibfnamefont{M.}~\bibnamefont{{Reuter}}},
  \bibinfo{journal}{Journal of High Energy Physics}
  \textbf{\bibinfo{volume}{10}}, \bibinfo{eid}{050} (\bibinfo{year}{2005}),
  \eprint{hep-th/0508202}.

\bibitem[{\citenamefont{Stoica}(2014)}]{Stoica2014Metric}
\bibinfo{author}{\bibfnamefont{O.~C.} \bibnamefont{Stoica}},
  \bibinfo{journal}{Annals of Physics} \textbf{\bibinfo{volume}{347}},
  \bibinfo{pages}{74} (\bibinfo{year}{2014}).

\bibitem[{\citenamefont{Svozil}(1987)}]{Svozil1987Quantum}
\bibinfo{author}{\bibfnamefont{K.}~\bibnamefont{Svozil}},
  \bibinfo{journal}{Journal of Physics A General Physics}
  \textbf{\bibinfo{volume}{20}}, \bibinfo{pages}{3861} (\bibinfo{year}{1987}).

\bibitem[{\citenamefont{Kroger}(2000)}]{H2000Fractal}
\bibinfo{author}{\bibfnamefont{H.}~\bibnamefont{Kroger}},
  \bibinfo{journal}{Physics Reports} \textbf{\bibinfo{volume}{323}},
  \bibinfo{pages}{81} (\bibinfo{year}{2000}).

\bibitem[{\citenamefont{Brezin et~al.}(1976)\citenamefont{Brezin, Zinn-Justin,
  and Le~Guillou}}]{brezin1976renormalization}
\bibinfo{author}{\bibfnamefont{E.}~\bibnamefont{Brezin}},
  \bibinfo{author}{\bibfnamefont{J.}~\bibnamefont{Zinn-Justin}},
  \bibnamefont{and}
  \bibinfo{author}{\bibfnamefont{J.}~\bibnamefont{Le~Guillou}},
  \bibinfo{journal}{Physical Review D} \textbf{\bibinfo{volume}{14}},
  \bibinfo{pages}{2615} (\bibinfo{year}{1976}).

\bibitem[{\citenamefont{Friedan}(1980{\natexlab{a}})}]{friedan1980nonlinear}
\bibinfo{author}{\bibfnamefont{D.}~\bibnamefont{Friedan}},
  \bibinfo{journal}{Physical Review Letters} \textbf{\bibinfo{volume}{45}},
  \bibinfo{pages}{1057} (\bibinfo{year}{1980}{\natexlab{a}}).

\bibitem[{\citenamefont{Friedan}(1980{\natexlab{b}})}]{Friedan1980}
\bibinfo{author}{\bibfnamefont{D.}~\bibnamefont{Friedan}},
  \bibinfo{journal}{Annals of Physics} \textbf{\bibinfo{volume}{163}},
  \bibinfo{pages}{318} (\bibinfo{year}{1980}{\natexlab{b}}).

\bibitem[{\citenamefont{Weinberg}(2008)}]{Weinberg2008}
\bibinfo{author}{\bibfnamefont{S.}~\bibnamefont{Weinberg}},
  \emph{\bibinfo{title}{Cosmology}} (\bibinfo{publisher}{Oxford University
  Press, Oxford}, \bibinfo{year}{2008}).

\bibitem[{\citenamefont{Riess et~al.}(1998)}]{Riess:1998cb}
\bibinfo{author}{\bibfnamefont{A.~G.} \bibnamefont{Riess}} \bibnamefont{et~al.}
  (\bibinfo{collaboration}{Supernova Search Team}),
  \bibinfo{journal}{Astron.J.} \textbf{\bibinfo{volume}{116}},
  \bibinfo{pages}{1009} (\bibinfo{year}{1998}), \eprint{astro-ph/9805201}.

\bibitem[{\citenamefont{Perlmutter et~al.}(1999)}]{Perlmutter:1998np}
\bibinfo{author}{\bibfnamefont{S.}~\bibnamefont{Perlmutter}}
  \bibnamefont{et~al.} (\bibinfo{collaboration}{Supernova Cosmology Project}),
  \bibinfo{journal}{Astrophys.J.} \textbf{\bibinfo{volume}{517}},
  \bibinfo{pages}{565} (\bibinfo{year}{1999}), \eprint{astro-ph/9812133}.

\bibitem[{\citenamefont{{Polchinski}}(2006)}]{2006hep.th....3249P}
\bibinfo{author}{\bibfnamefont{J.}~\bibnamefont{{Polchinski}}},
  \bibinfo{journal}{ArXiv High Energy Physics - Theory e-prints}
  (\bibinfo{year}{2006}), \eprint{hep-th/0603249}.

\bibitem[{\citenamefont{{Mass{\'o}}}(2009)}]{2009PhLB..679..433M}
\bibinfo{author}{\bibfnamefont{E.}~\bibnamefont{{Mass{\'o}}}},
  \bibinfo{journal}{Physics Letters B} \textbf{\bibinfo{volume}{679}},
  \bibinfo{pages}{433} (\bibinfo{year}{2009}), \eprint{0902.4318}.

\bibitem[{\citenamefont{Padmanabhan and Singh}(1987)}]{padmanabhan1987response}
\bibinfo{author}{\bibfnamefont{T.}~\bibnamefont{Padmanabhan}} \bibnamefont{and}
  \bibinfo{author}{\bibfnamefont{T.}~\bibnamefont{Singh}},
  \bibinfo{journal}{Classical and Quantum Gravity}
  \textbf{\bibinfo{volume}{4}}, \bibinfo{pages}{1397} (\bibinfo{year}{1987}).

\bibitem[{\citenamefont{{Padmanabhan}}(2005)}]{2005CQGra..22L.107P}
\bibinfo{author}{\bibfnamefont{T.}~\bibnamefont{{Padmanabhan}}},
  \bibinfo{journal}{Classical and Quantum Gravity}
  \textbf{\bibinfo{volume}{22}}, \bibinfo{pages}{L107} (\bibinfo{year}{2005}),
  \eprint{hep-th/0406060}.

\bibitem[{\citenamefont{Jaffe}(2005)}]{PhysRevD.72.021301}
\bibinfo{author}{\bibfnamefont{R.~L.} \bibnamefont{Jaffe}},
  \bibinfo{journal}{Phys. Rev. D} \textbf{\bibinfo{volume}{72}},
  \bibinfo{pages}{021301} (\bibinfo{year}{2005}).

\bibitem[{\citenamefont{Chow and Knopf}(2004)}]{chow2004ricci}
\bibinfo{author}{\bibfnamefont{B.}~\bibnamefont{Chow}} \bibnamefont{and}
  \bibinfo{author}{\bibfnamefont{D.}~\bibnamefont{Knopf}},
  \emph{\bibinfo{title}{The Ricci flow: an introduction}},
  vol.~\bibinfo{volume}{1} (\bibinfo{publisher}{American Mathematical Soc.},
  \bibinfo{year}{2004}).

\end{thebibliography}

\end{acknowledgments}

\end{document}